\documentclass[onecolumn]{article}
\usepackage[paperwidth=8.5in, paperheight=11in]
    {geometry}
\usepackage{fullpage}
\usepackage[latin9]{inputenc}
\usepackage{changepage}
\usepackage{amsmath}
\usepackage{amssymb}
\usepackage{graphicx}
\usepackage{morefloats}
\usepackage{verbatim}
\usepackage{pdflscape}
\usepackage{ifthen}
\usepackage{verbatim}
\usepackage[justification=centering]{caption}
\usepackage{subfig}
\usepackage[colorlinks=false]{hyperref}
\usepackage[all]{hypcap}

\providecommand{\tabularnewline}{\\}
\usepackage{fullpage}\usepackage[justification=centering]{caption}\usepackage[all]{hypcap}\usepackage{microtype}
\def\newsqrt{\mathpalette\DHLhksqrt}
\def\DHLhksqrt#1#2{%
\setbox0=\hbox{$#1\sqrt{#2\,}$}\dimen0=\ht0
\advance\dimen0-0.2\ht0
\setbox2=\hbox{\vrule height\ht0 depth -\dimen0}%
{\box0\lower0.4pt\box2}}
\begin{document}
\title{Analytical and Numerical Calculations for the Asymptotic Behaviors
of Unitary $9j$ Coefficients}
\author{Brian Kleszyk, Larry Zamick \\
 Department of Physics and Astronomy, Rutgers University, Piscataway,
New Jersey 08854 \\
 }
\date{\today}
\maketitle

\begin{abstract}
Previously it was noted in numerical calculations that a certain unitary
$9j$ coefficient 
\[
U(I,j)={\Big<(jj)^{2j}(jj)^{2j}\Big|(jj)^{2j}(jj)^{(2j-2)}\Big>}^{I}
\]
decreases with increasing $j$ and for fixed small $I$. The decrease
is of the form $Aj^{m}e^{-\alpha j}$. The exponential decay factor
dominates. Analytically we also show using the Stirling approximation,
that $\alpha=4\ln(2)$ and $m=\frac{3}{2}$.
\end{abstract}

\section{Introduction}

In previous works \cite{Zamick1,Zamick2} Zamick and Escuderos addressed
the problem of maximum $j$-pairing. In the course of these studies
they found that results simplified by the fact that a certain coupling
matrix element was very small. This was the unitary $9j$ coefficient
\begin{equation}
U(I,j)={\Big<(jj)^{2j}(jj)^{2j}\Big|(jj)^{2j}(jj)^{(2j-2)}\Big>}^{I}
\end{equation}
for small $I$ e.g. $I=2$. The work started in the $g_{9/2}$ shell.
But as one went to higher shells this $U9j$ becomes rapidly smaller.
Indeed behavior was parametrized as $Aj^{m}e^{-\alpha j}$ \cite{Zamick2,Kleszyk}.
The consequence of a very weak coupling is that for small total angular
momentum $I$ the lowest 2 states for a maximum $J$ pairing interaction
are ${\Big<(jj)^{2j}(jj)^{2j}\Big|(jj)^{J_{p}}(jj)^{J_{n}}\Big>}^{I}$
and ${\Big<(jj)^{2j}(jj)^{(2j-2)}\Big|(jj)^{J_{p}}(jj)^{J_{n}}\Big>}^{I}$
with $J_{p}$ and $J_{n}$ both even \cite{Zamick1,Zamick2}. In this
work we will first conduct numerical studies to much higher angular
momenta and with greater precision for the unitary $9j$ coefficients
using Mathematica. We will then approach the problem analytically
and derive the parameters $\alpha$ and $m$. We also consider cases
where $I$ is large.


\section{Calculation}

\subsection{Asymptotes of Small $I$}

As was noted in \cite{Zamick1} at first glance $U(2,j)$ seems to
fall of exponentially with $j$. This suggests a form 
\begin{equation}
Ae^{-\alpha j}
\end{equation}
For this form $\ln(|U(2,j)|)=\ln(A)-\alpha j$. If this were true
there would be a linear relationship between $\ln(|U(2,j)|)$ and
$j$. We will here also consider other values of $I$ as indicated
above.

\begin{figure}[h!]
    \centering
	\includegraphics[width=.7\textwidth]
	  {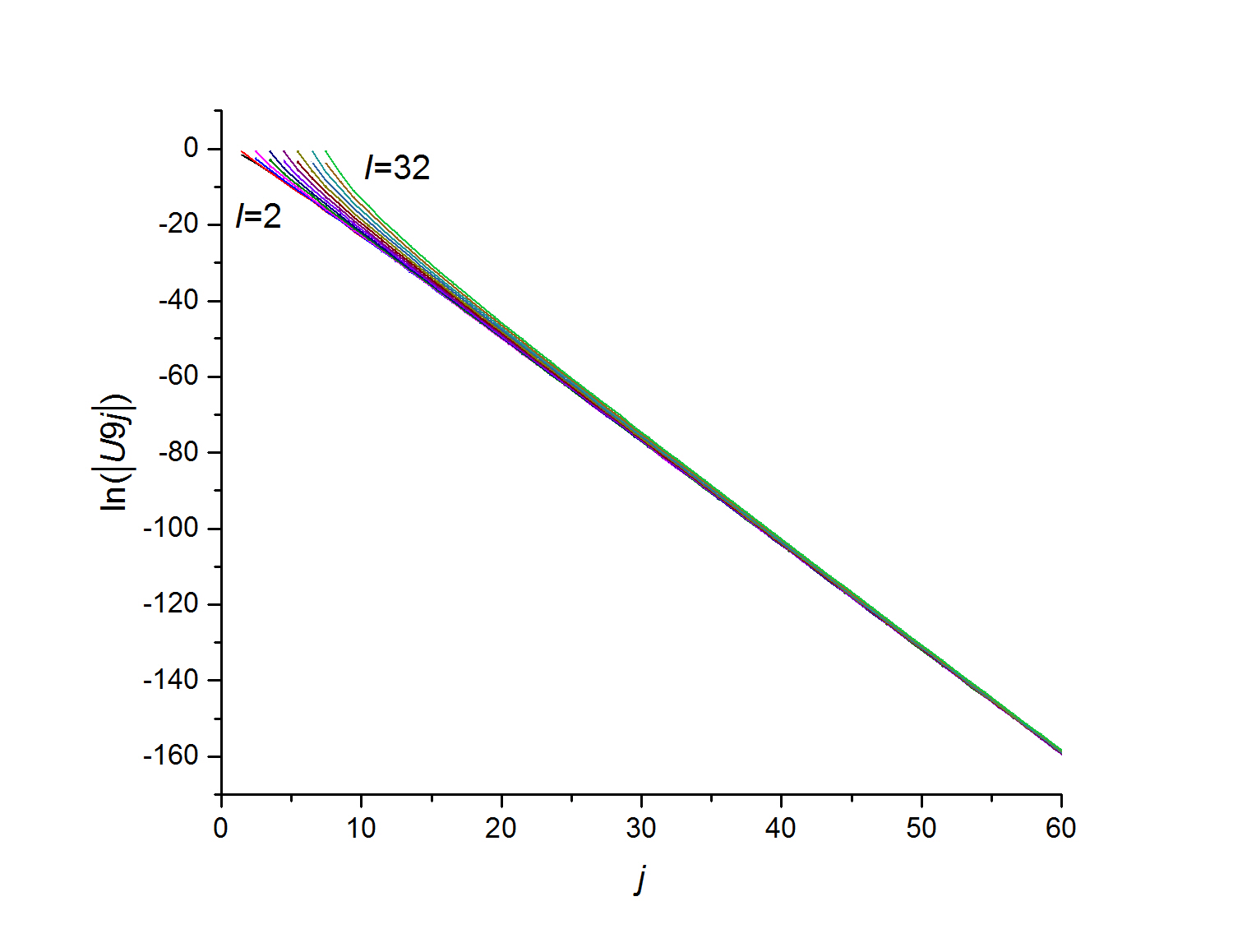}
	\caption{(color online)$\ln(|U9j|)$ vs $j$ \\
		$I=2,4,6,...,32$
		\label{fig:lngraph} 
	} 
\end{figure}

We first plot, in Figure~\ref{fig:lngraph}, $\ln(|U(I,j)|)$ vs
$j$ for all even $I$ values between $I=2$ and $I=32$. The curves
indeed approach straight lines indicating that the $U(I,j)$'s drops
exponentially with $j$. This is certainly the dominant trend but
there are small deviations indicated by the error analysis.

\begin{figure}[h!]
    \centering
	\parbox{0.47\textwidth}{ \centering
		\includegraphics[width=.5\textwidth]
		  {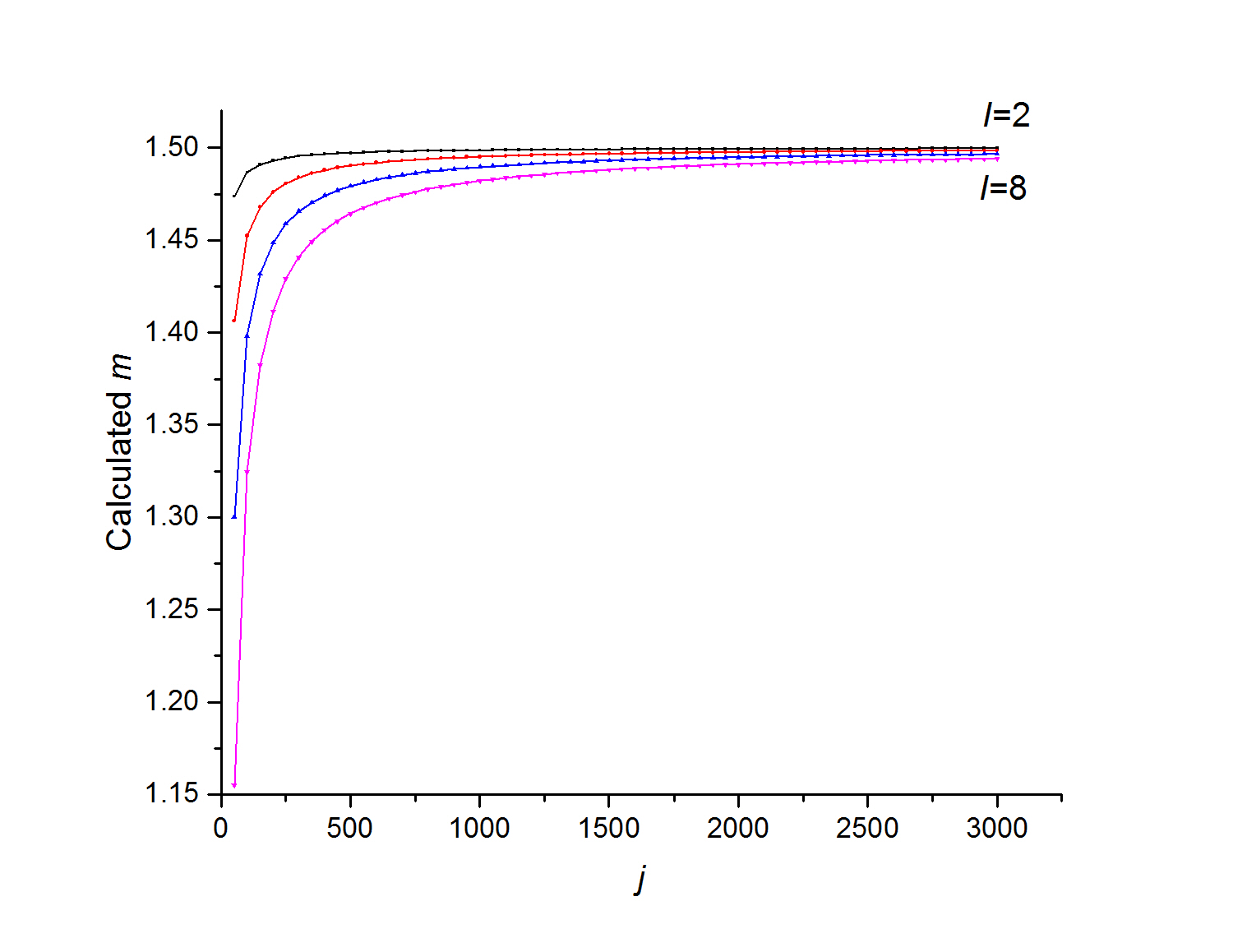}
		\caption{(color online)Suspected $m$ vs $j$ \\ $I=2,4,6,8$
			\label{fig:firstmgraph}
		} 
	} 
	\qquad
	\parbox{0.47\textwidth}{ \centering
		\includegraphics[width=.5\textwidth]
		  {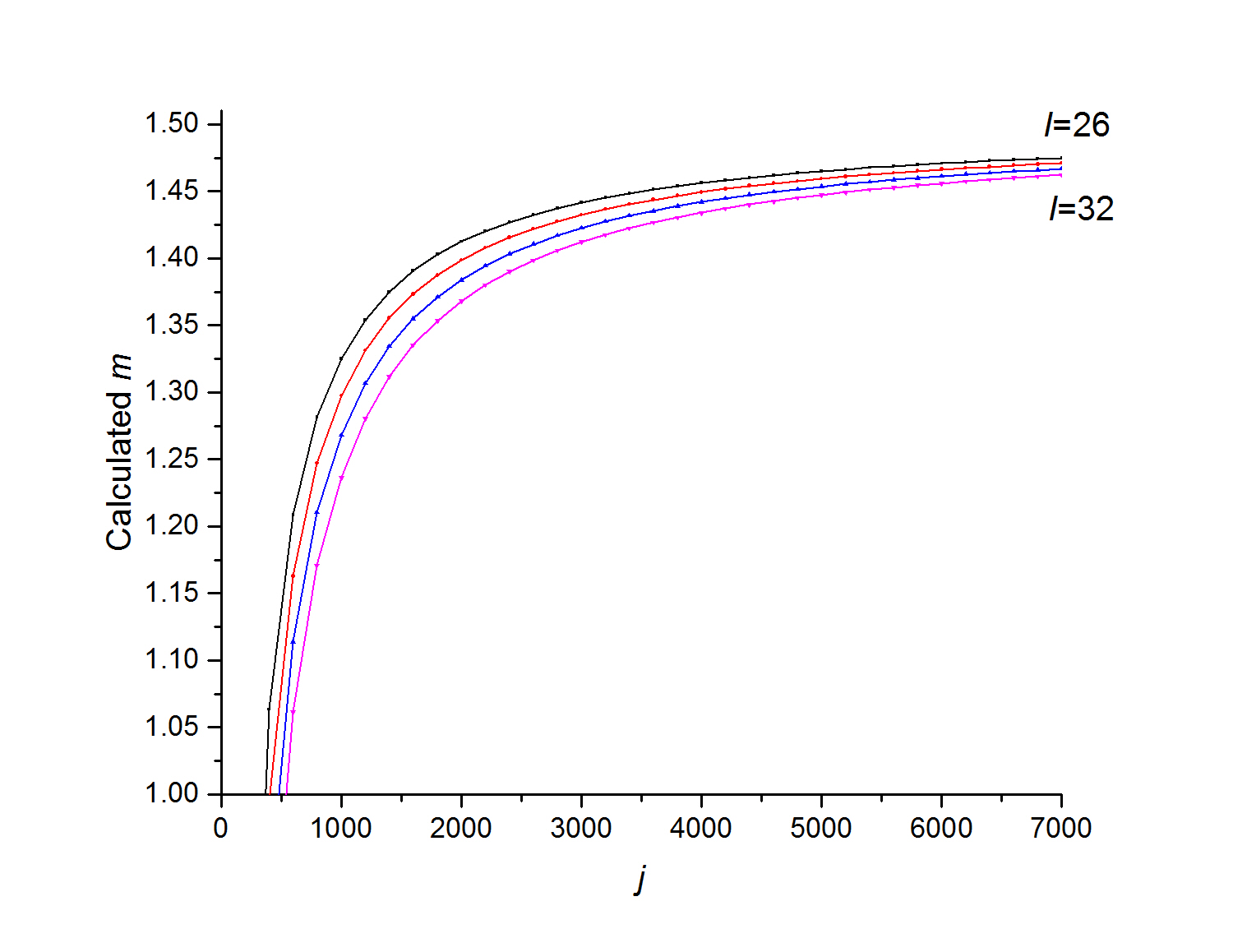}
		\caption{(color online)Suspected $m$ vs $j$ \\ $I=26,28,30,32$
			\label{fig:lastmgraph} 
		} 
	}
\end{figure}

\begin{figure}[h!]
     \centering
	\includegraphics[width=.7\textwidth]
	 {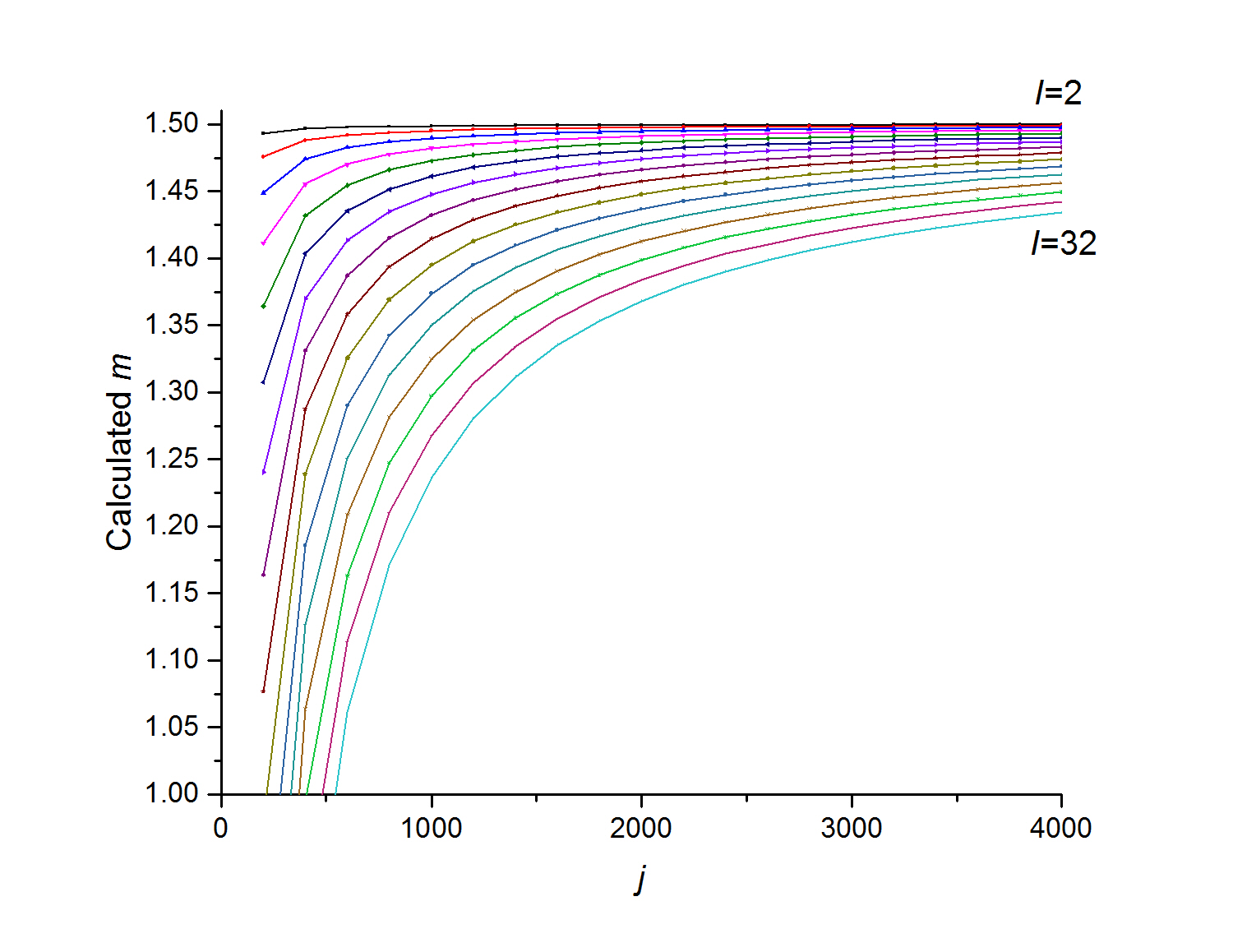}
		\caption{(color online)Suspected $m$ vs $j$ \\ $I=2,4,6,...,32$
			\label{fig:allmgraph}
		} 
\end{figure}

We try a more elaborate form 
\begin{equation}
UA(I,j)=Aj^{m}e^{-\alpha j}\label{eq:asymp}
\end{equation}
We consider the ratio 
\begin{equation}
RR=\dfrac{\text{U}(I,j+1)^{2}}{\text{U}(I,j)\text{\space}\text{U}(I,j+2)}
\end{equation}
If we assume that $U9j=Aj^{m}e^{-\alpha j}$, then we have 
\begin{equation}
RR=\dfrac{(A(j+1)^{m}e^{-\alpha(j+1)})^{2}}{Aj^{m}e^{-\alpha j}\times A(j+2)^{m}e^{-\alpha(j+2)}}
\end{equation}
With some algebra this becomes 
\begin{equation}
=\dfrac{e^{-2\alpha j}e^{-2\alpha}(j+1)^{2m}}{e^{-2\alpha j}e^{-2\alpha}j^{m}(j+2)^{m}}
\end{equation}
It is obvious to see the factors which cancel out, then we take the
ln of both sides and obtain 
\begin{equation}
\ln(RR)=m\ln\left(\dfrac{(j+1)^{2}}{j(j+2)}\right)
\end{equation}
We therefor have the extracted $m$ 
\begin{equation}
m=\dfrac{\ln(RR)}{\ln\left(\dfrac{(j+1)^{2}}{j(j+2)}\right)}
\end{equation}
It should be noted that in the large $j$ limit $\dfrac{(j+1)^{2m}}{(j(j+2))^{m}}$
approaches $1+\dfrac{m}{j^{2}}$. We plot some cases of $m$ vs. $j$
in the attached Figures~\ref{fig:firstmgraph} to~\ref{fig:allmgraph}.
We find that all even $I$ from $I=2$ to $I=12$, $m$ converges
to 1.5 in the large $j$ limit.

It is important to note that in order to obtain the asymptotic value
of $m$ in Eq.(\ref{eq:asymp}) one must go to a sufficiently large
value of $j$. Furthermore the bigger the value of $I$ the higher
one has to go in $j$. To show the perils of choosing the maximum
$j$ too small suppose we choose it to by 500.5, which a priori most
would consider to be a very large number. The values of $m$ for $I=2,4,10,20,30$
respectively 1.495, 1.481, 1.391, 1.085, and 0.577. We now see a steady
decrease in $m$ as $I$ increases, which could lead to the false
conclusion that there is a different asymptotic value of $m$ for
each $I$. However when we choose $j$ large enough e.g. up to 7000.5
for $I=32$ we see that the asymptotic value of $m$ is the same for
all even $I$ up to $I=32$, namely $m=1.5$. It should be noted that
convergence is slower as $I$ increases. 

\subsection{Asymptotes of Large $I$}

\begin{figure}[h!]
    \centering
	\parbox{0.47\textwidth}{ \centering
		\includegraphics[width=.5\textwidth]
		  {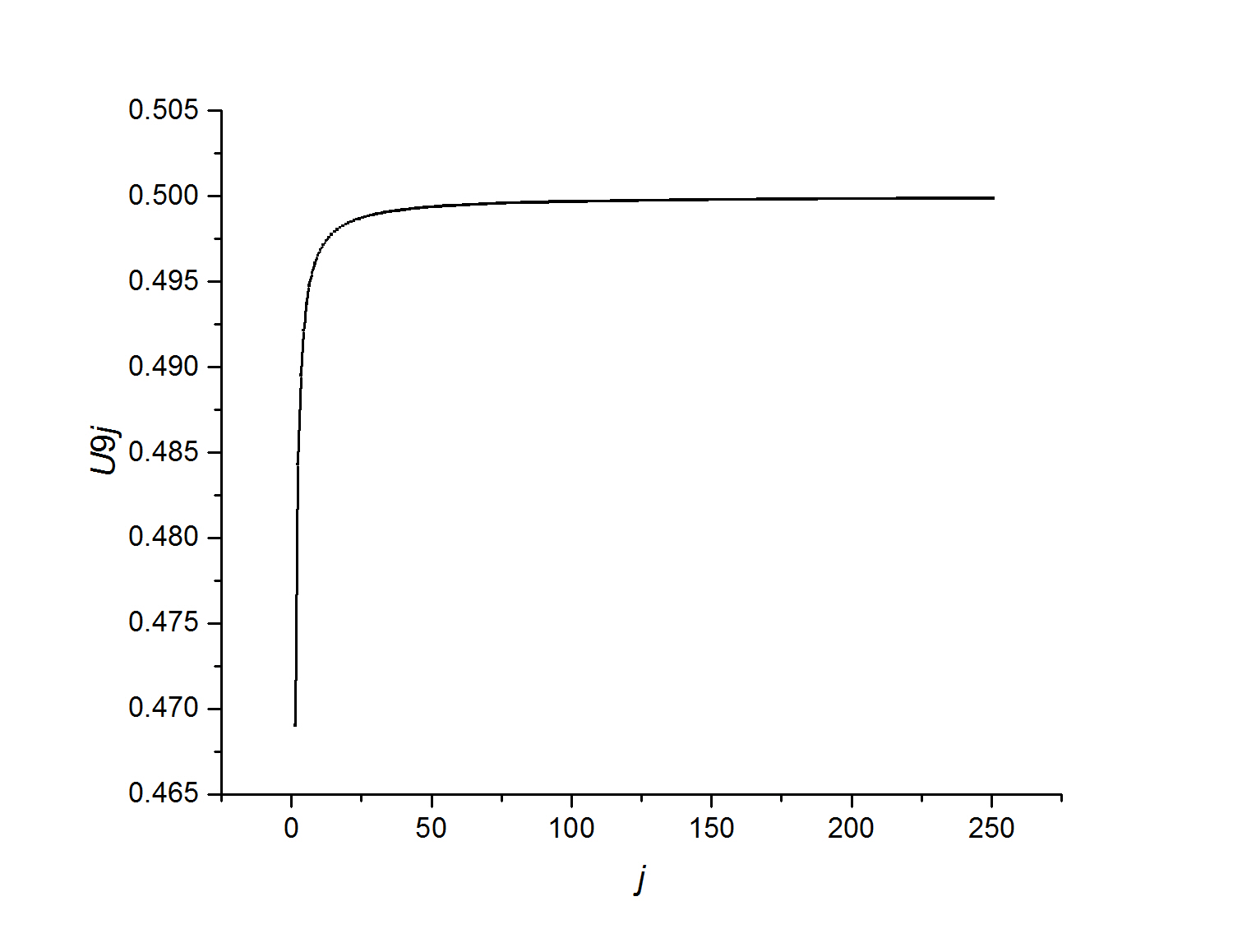}
		\caption{(color online)$U9j$ vs $j$,
			$I=I_{\text{max}}$\\
			$(n=0)$
			\label{fig:justImax} 
		} 
	} 
	\qquad
	\parbox{0.47\textwidth}{ \centering
		\includegraphics[width=.5\textwidth]
		  {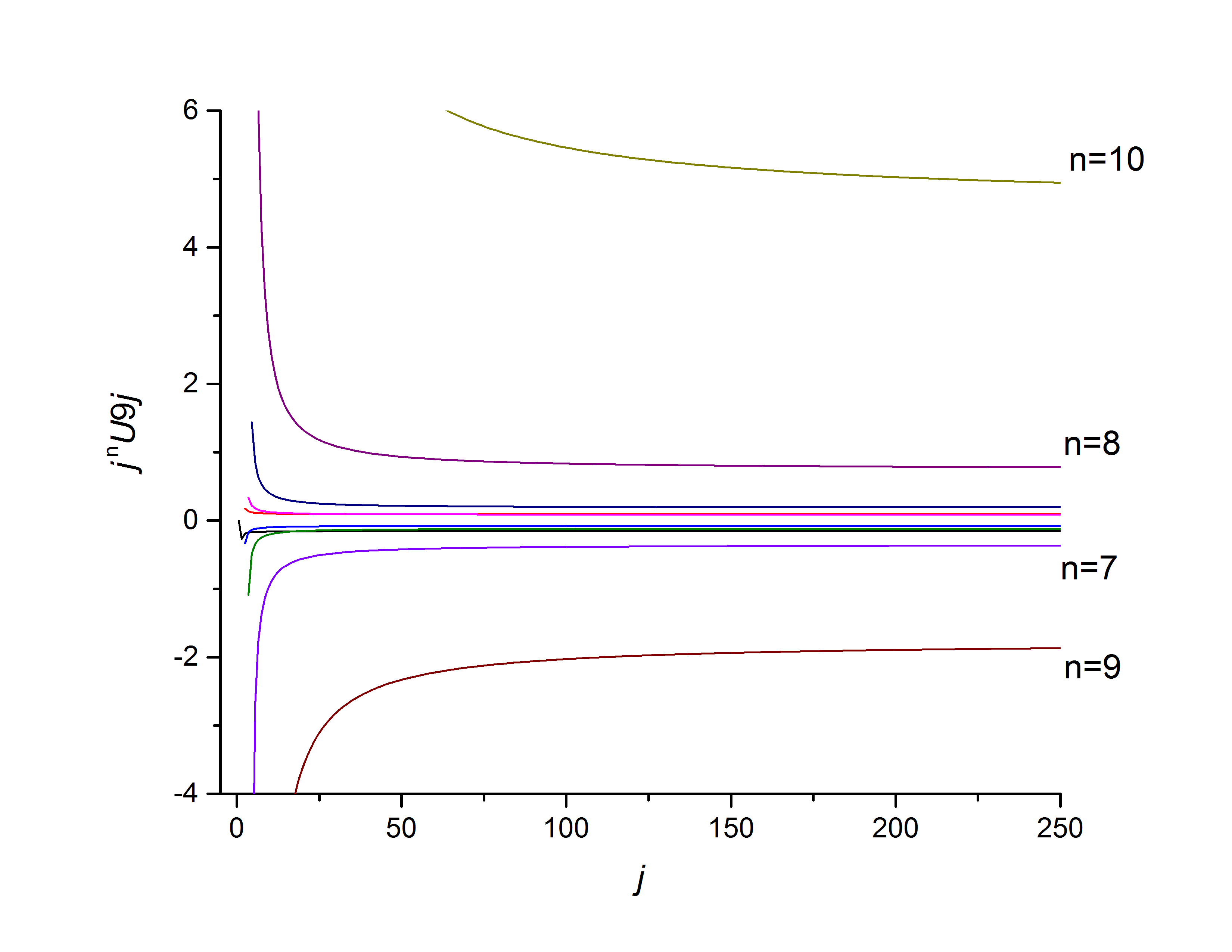}
		\caption{(color online)$(j^n U9j)$ vs $j$,
			$I=I_{\text{max}}-2n$ \\
			$n=1,2,...,10$
			\label{fig:allImax} 
		} 
	}
\end{figure}

We next consider $U(I,j)$ for the largest values of $I$. We start
with $I=I_{\text{max}}=4j-2$ and then also consider $I_{\text{max}}-2$,
$I_{\text{max}}-4$, etc. We find that $U(I_{\text{max}},j)$ approaches
a constant for large $j$ shown in Figure~\ref{fig:justImax}. We
assume that the form of the asymptote is 
\begin{equation}
	U(I_{\text{max}}-2n,j)=\dfrac{A}{j^{n}}\label{eq:maxasymp}
\end{equation}
Then we plot $U(I_{\text{max}}-2n,j)\times j^{n}$ versus $j$ to
determine if this value approaches a constant. The results are shown
in Figure~\ref{fig:allImax}. We can conclude then that the asymptote
for large $I$ adheres to Eq.(\ref{eq:maxasymp}).

A formula involving many factorials for the case $I=I_{\text{max}}$
is also given by Varshalovich et al. in sec.10:8:4 Eq.(14) in \cite{Varshalovich}.
We finally remind the reader that our motivation for this work comes
from our desire to ether understand the wave function arising from
a {}``maximum $J$-pairing'' Hamiltonian \cite{Zamick1,Zamick2}.

\section{Analytical Results}

\subsection{Asymptotes of Small $I$}

The numerical results in the previous section for the small $I$ cases
lead to the result $m=1.5$ and the figures showed a dominantly exponential
decrease with $j$ \cite{Kleszyk}. We can show some analytical results.
We note that there is an explicit formula for the $9j$ symbol associated
with the unitary $9j$ coefficient above in the work of Varshalovich
et al. \cite{Varshalovich} sec 10:8:3 Eq.(9) shown here: 
\[
9j=\left\{ \begin{array}{ccc}
a & b & c\\
d & e & f\\
a+d & b+e & j
\end{array}\right\} =\left<cf\text{\space}(a-b)(d-e)|j(a-b+d-e)\right>
\]
\begin{equation}
\times\left[\dfrac{(2a)!(2b)!(2d)!(2e)!(a+b+d+e+j+1)!(a+d+e+b-j)!}{(2a+2d+1)!(2b+2e+1)!(a+b+c+1)!(a+b-c)!(d+e+f+1)!(d+e-f)!(2j+1)}\right]^{\frac{1}{2}}
\end{equation}
We associate $a,b,d,e\rightarrow j$; $c,(a+d),(b+e)\rightarrow2j$;
$f\rightarrow(2j-2)$; and $j\rightarrow I$. For some simplification
we define a new variable $J=2j$. We apply that expression to this
problem and consider $U9j$ rather than $9j$, 
\begin{equation}
U(I,j)=\dfrac{(J!)^{2}}{(2J)!}\left[\dfrac{(2J+I+1)!(2J-I)!(2J+1)(2J-3)}{(2J+1)!(2J-1)!}\right]^{\frac{1}{2}}\times\sqrt{\dfrac{1}{2(2I+1)}}\times\left<J(J-2)00|I0\right>
\end{equation}
Thus we have related the $U9j$ to a Clebsch-Gordan(CG) coefficient.

For the particular $U9j$ above and for $I=2$ we obtain the following
expression 
\begin{equation}
U(2,j)=\dfrac{(J!)^{2}}{(2J)!}\left[\dfrac{(2J+1)(2J+3)(2J+2)(2J-3)}{(2J-1)}\right]^{\frac{1}{2}}\times\sqrt{\dfrac{1}{10}}\times\left<J(J-2)00|20\right>
\end{equation}
This special $U9j$ is proportional to a Clebsch-Gordan coefficient.
There is a useful formula in Talmi's book \cite{Talmi} for the associated
$3j$ symbol shown here: 
\begin{equation}
\left(\begin{array}{ccc}
j_{1} & j_{2} & j_{3}\\
0 & 0 & 0
\end{array}\right)=\dfrac{1}{2}\left(1+(-1)^{j_{1}+j_{2}+j_{3}}\right)(-1)^{g}\times\sqrt{\dfrac{(2g-2j_{1})!(2g-2j_{2})!(2g-2j_{3})!}{(2g+1)!}}\times\dfrac{g!}{(g-j_{1})!(g-j_{2})!(g-j_{3})!}
\end{equation}
where $2g=j_{1}+j_{2}+j_{3}$ and 
\begin{equation}
CG=\newsqrt{(2j_{3}+1)}(-1)^{j_{1}-j_{2}}\left(\begin{array}{ccc}
j_{1} & j_{2} & j_{3}\\
0 & 0 & 0
\end{array}\right)\label{eq:clebsch}
\end{equation}

There is a simpler formula in Talmi's book \cite{Talmi} for this
coefficient when $I=2$: 
\begin{equation}
\left<J(J-2)00|20\right>=-\sqrt{\dfrac{15J(J-1)^{2}}{((2J-3)(2J-2)(2J-1)(2J+1))}}
\end{equation}
It is easy to see that the CG coefficient falls off as $\dfrac{1}{\sqrt{J}}$.
We now get the combined expression 
\begin{equation}
U(2,j)=\dfrac{(J!)^{2}}{(2J)!}\sqrt{\dfrac{3J(J-1)^{2}(2J+3)(2J+2)}{2(2J-2)(2J-1)^{2}}}
\end{equation}
The exponential behavior comes from the factorials via the Stirling
approximation 
\begin{equation}
\ln(n!)\approx n\ln(n)-n
\end{equation}
If we stop there we get 
\begin{equation}
\ln\left(\dfrac{(J!)^{2}}{(2J)!}\right)\approx2J\ln(J)-2J\ln(2J)=-2J\ln2
\end{equation}
However to get the correct asymptotic behavior we must go beyond this
and include one more term to obtain the more accurate Stirling approximation
\begin{equation}
\ln(n!)=n\ln(n)-n+\ln(\sqrt{2\pi n})
\end{equation}
Using the extended Stirling approximation this becomes 
\begin{equation}
\ln\left(\dfrac{(J!)^{2}}{(2J)!}\right)\approx-2J\ln(2)+\ln(\sqrt{\pi J})
\end{equation}
Recall that we had assigned $J=2j$ and then taking a inverse logarithm
of this yields a contribution 
\begin{equation}
\dfrac{(J!)^{2}}{(2J)!}\approx e^{-4\ln(2)j}\sqrt{2\pi j}
\end{equation}
When we go from $j$ to $j+1$ we get a decreases of about 16 from
the exponential factor. This decrease dominates over the increase
from the second factor. The second factor and the other terms must
contribute to get the $j^{m}$ part which serves to reduce this ratio
a bit.

If the {}``small'' term in the Stirling approximation is neglected
a problem arises. The factors under the square root sign clearly go
as $j^{3/2}$ in the large $j$ limit. However the Clebsch-Gordan
coefficient decreases with $j$. This leads to an effective $m$ less
than $\frac{3}{2}$. However numerical calculations \cite{Kleszyk}
clearly indicate that $m=\frac{3}{2}$. Hence, although the simplest
version of the Stirling approximation gives the right exponential
behavior it gives the wrong $j^{m}$ dependence. By including the
{}``small correction'' we take care of this problem.

Analytic expressions of specific $9j$ coefficients have been previously
considered for special cases e.g. for the case of partial dynamical
symmetries by Robinson and Zamick \cite{Robinson}. Many relations
for $9j$ symbols were found by Zhao and Arima \cite{Zhao} in the
context of maximum $j$-pairing hamiltonians. Explicit studies of
the asymptotic behavious of $9j$ coefficients have been performed
by Anderson et. al. \cite{Anderson} and by Yu and Littlejohn \cite{Yu}.
What distinguishes the present work from the ones just mentioned is
that only here do we consider $9j$s which display an exponential
decrease with increasing $j$. This is called non-classical behavior
by the experts. The large difference in behavior comes from the fact
that we are considering coupling matrix elements involving 2 different
$J$ values $2j$ and $2j-2$ whereas in Zhao and Arima \cite{Zhao}
for the problem they are addressing they have the same $J$ values.
Ironically we have to be in the non-classical region mathematically
to reach the classical limit for the physical problem in question.

\subsection{Asymptotes of Large $I$}

We now consider the region near $I=I_{\text{max}}=4j-2$. It should
be pointed out that whereas in the small $I$ case we kept $I$ fixed
as we increased $j$, here as we change $j$ we change $I$. Thus
we are making different comparisons. The figures confirm that for
this analysis there is a power law behavior rather than an exponential
one. The $U9j$ goes as $\dfrac{1}{j^{n}}$ where $n=\dfrac{(I_{\text{max}}-I)}{2}$.
It should be noted that for $I=I_{\text{max}}=4j-2$ the value of
the $U9j$ was shown by Talmi~\cite{Talmipri} to be 
\begin{equation}
U9j=\dfrac{\sqrt{(2j-1)(8j-1)}}{(8j-2)}
\end{equation}

Note that this 9j approaches 1/2 when j becomes very large. Subsequently
an alternate proof was provided by Bayman\cite{Bayman}.

For large $I$ we write $I=4j-2-2n$ and assume $n$ is much smalller
than $j$. We use a more general formula in Talmi's book \cite{Talmi}
(top of page 960) for $\big<2j(2j-2)00\big|I0\big>$.

We can get an expression for all $n$ by using the Stirling approximation
for factorials involving large parameters but not for those involving
only $n$. We obtain the following result: 
\begin{equation}
U9j=\dfrac{(-1)^{n}}{2\newsqrt{2}16^{n}}\dfrac{\sqrt{\left((2n+2)!(2n)!\right)}}{(n!)j^{n}}
\end{equation}
as $j$ becomes very large. One can verify that for $n=0$ this is
indeed $\dfrac{1}{2}$ and note that for $n=1$ we get $\dfrac{\sqrt{3/2}}{8j}$.
One notes that in this limit ($n$ much smaller than $j$) the Clebsch-Gordan
coefficient goes as $\dfrac{1}{j^{1/4}}$ (alternatively the $3j$
goes as $\dfrac{1}{j^{3/4}}$) in the large $j$ limit. In more detail
we have: 
\begin{equation}
CG=\dfrac{\sqrt{(2n)!}}{n!(2^{n})}\left(\dfrac{1}{\pi j}\right)^{1/4}(-1)^{n}
\end{equation}
The $\dfrac{1}{j^{1/4}}$ behavior in the large $I$ limit is in contrast
to the behavior in the previous section where $I$ was fixed at a
small value whilst $j$ was increased. In that case the Clebsch-Gordan
coefficient from~Eq.(\ref{eq:clebsch}) went as $\dfrac{1}{j^{1/2}}$.
In this work our motivation for studying the specific $U9j$ coefficients
above, was to better understand the wave functions of a maximum $J$-pairing
hamiltonian. What we had previously shown numerically we now have
attempted to show analytically. We found the numerical results crucial
in guiding us to the analytical ones. We have succeeded in getting
analytic expressions for the asymptotic behaviors for small $I$ by
using the extended Stirling approximation. We are also able to make
statements about the large $I$ problem.

We would like to thank Ben Bayman and Igal Talmi for their valuable
help and interest. 

Brian Kleszyk also thanks the Rutgers Aresty Research Center for undergraduate
research for support during the 2013-2014 academic year.
\appendix
\section{Appendix}

In this paper we focus on equations (11 and 13), and (23 and 24) of
the work of Kleszyk and Zamick \cite{Kleszyk}. In particular we consider
the case when the total angular momentum $I$is equal to $I_{\text{max}}-2n$
and $I_{\text{max}}\equiv4j-2$, and $n=0,1,2,...$ We take the limit
of large $j$ where $n$ becomes much smaller than $j$. We also define
$J=2j$, where $j$ is the .
 We first address the 3$j$ coefficient: 
 
\begin{equation}
\left(\begin{array}{ccc}
2J & 2J-2 & I\\
0 & 0 & 0
\end{array}\right)
\end{equation}
We then also can define $I$ using a new variable $m$ with $I=4j-2m$,
and this time $m=1,2,3,...$ We can separate parts of the 3$j$ which
now becomes 
\begin{equation}
3j=\dfrac{\sqrt{(2m-1)!}}{(m-1)!}(-1)^{m}\sqrt{\dfrac{N_{1}!N_{2}!}{N_{3}!}}\dfrac{N_{4}!}{N_{5}!N_{6}!}
\end{equation}
where the 6 factors $N_{i}$ are: 
\begin{equation}
\tag{2a}\begin{array}{ccc}
N_{1}=2J-2-2m & N_{2}=2J+2-2m & N_{3}=4J-1-2m\\
N_{4}=2J-1-m & N_{5}=J-1-m & N_{6}=J+1-m
\end{array}\label{eq:defN}
\end{equation}
We use the Stirling approximation, 
 
\begin{equation}
\ln N!=N\ln N-N+\ln\sqrt{2\pi N}
\end{equation}
and it should be noted that the approximation approaches the true
value asymptotically. Now we can write: $N_{i}=(\alpha_{i}+\beta_{i}m+\gamma_{i}J)$
with differing constant coefficients. In Eq.\eqref{eq:defN} we give
the contribuition of $-N,\ln\sqrt{2\pi N},\alpha\ln N,m\beta\ln N,$
and $\gamma J\ln N$. For the latter we break things up into (a)``extreme''
and (b) ``next order''. This is necessary because ``next order''
has contriutions comparable to those in ``$-N$''. 

\begin{table}[h]
\centering \caption{Asymptotic contributions to the 3j coefficients}

\label{table:3j} \subfloat[]{%
\begin{tabular}{l|r|r|r}
 & $-N_{i}$  & $\ln\sqrt{2\pi N_{i}}$  & $\alpha_{i}\ln N_{i}$ \tabularnewline
\hline 
(1)  & $\dfrac{}{}-\frac{1}{2}(2J-2-2m)$  & $\frac{1}{2}\ln\sqrt{4\pi J}$  & $-\ln(2J)$ \tabularnewline
(2)  & $\dfrac{}{}-\frac{1}{2}(2J+2-2m)$  & $\frac{1}{2}\ln\sqrt{4\pi J}$  & $-\ln(2J)$ \tabularnewline
(3)  & $\dfrac{}{}\frac{1}{2}(4J-1-2m)$  & $\frac{1}{2}\ln\sqrt{4\pi J}$  & $-\ln(2J)$ \tabularnewline
(4)  & $\dfrac{}{}\frac{}{}-(2J-1-m)$  & $\frac{1}{2}\ln\sqrt{4\pi J}$  & $-\ln(2J)$ \tabularnewline
(5)  & $\dfrac{}{}(J-1+m)$  & $\frac{1}{2}\ln\sqrt{4\pi J}$  & $-\ln(2J)$ \tabularnewline
(6)  & $\dfrac{}{}(J+1-m)$  & $\frac{1}{2}\ln\sqrt{4\pi J}$  & $-\ln(2J)$ \tabularnewline
\hline 
Total  & $\dfrac{}{}\frac{1}{2}$  & $\ln\left(\frac{2}{\pi J}\right)^{1/4}$  & $\ln\frac{1}{\sqrt{J}}$ \tabularnewline
\end{tabular}

}

\hspace{1cm} \subfloat[]{%
\begin{tabular}{l|r|r|r}
 & $\beta_{i}m\ln N_{i}$  & $\gamma_{i}J\ln N_{i}$  & $\gamma_{i}J\ln N_{i}$ \tabularnewline
\hline 
(1)  & $\dfrac{}{}-m\ln(2J)$  & $J\ln J$  & $-1-m$ \tabularnewline
(2)  & $\dfrac{}{}-m\ln(2J)$  & $-J\ln(2J)$  & $1-m$ \tabularnewline
(3)  & $\dfrac{}{}m\ln(4J)$  & $-2J\ln(2J)$  & $\frac{1}{2}+m$ \tabularnewline
(4)  & $\dfrac{}{}-m\ln(2J)$  & $2J\ln(2J)$  & $-1-m$ \tabularnewline
(5)  & $\dfrac{}{}m\ln J$  & $-J\ln J$  & $1+m$ \tabularnewline
(6)  & $\dfrac{}{}m\ln J$  & $-J\ln J$  & $-1+m$ \tabularnewline
\hline 
Total  & $\dfrac{}{}-m\ln2$  & $0$  & $-\frac{1}{2}$ \tabularnewline
\end{tabular}

}
\end{table}

First notice that ``$\gamma J\ln N$'' result is $\dfrac{1}{2}$,
which cancels the $+\dfrac{1}{2}$ from ``$-N$''. Adding up all
the totals we get 
\begin{equation}
-m\ln2+\ln\left(\dfrac{2}{\pi J}\right)^{1/4}+\ln\left(\dfrac{1}{\sqrt{J}}\right)
\end{equation}
\begin{equation}
=-m\ln2+\ln\left(\dfrac{2}{\pi J^{3}}\right)^{1/4}
\end{equation}
Taking the antilog we get 
\begin{equation}
3j\approx e^{m\ln2}\left(\dfrac{2}{\pi J^{3}}\right)^{1/4}
\end{equation}
and note that $e^{-m\ln2}=\dfrac{1}{2^{m}}$.

Putting everything together and putting things in terms of $j$ and
$n$ we obtain 
\begin{equation}
3j\rightarrow\dfrac{\sqrt{(2n)!}}{n!2^{n}}(-1)^{n}\left(\dfrac{1}{64\pi j^{3}}\right)^{1/4}
\end{equation}
We see that in the limit $n\ll j$, $3j$ goes as $\dfrac{1}{j^{3/4}}$.
Alternatively the Clebsch-Gordan has an asymptotic value 
\begin{equation}
CG\rightarrow\dfrac{\sqrt{(2n)!}}{n!2^{n}}(-1)^{n}\left(\dfrac{1}{\pi j}\right)^{1/4}
\end{equation}

\subsection{The Unitary $9j$ coefficient}

Again we will write $I=4j-2m$, with $m=1,2,3,...$ and we can rewrite
Eq.(11) from \cite{Kleszyk} as 
\begin{equation}
U(9j)=\dfrac{FAC}{\sqrt{PROD}}\sqrt{\dfrac{(2J+1)(2J-3)}{2}}\times3j
\end{equation}
where 
\begin{equation}
FAC=\dfrac{(C_{1}!)^{2}}{C_{2}!}\sqrt{\dfrac{C_{3}!}{C_{4}!C_{5}!}}
\end{equation}
with 
\[
\begin{array}{rrrrr}
C_{1}=J & C_{2}=2J & C_{3}=4-J+1 & C_{4}=2J+1 & C_{5}=2J-1\end{array}
\]
Then we also have 
\begin{equation}
PROD=(4J+1)(4J)...(4J-m)
\end{equation}
There are $2m$ terms in $PROD$. Asymptotically we obtain 
\begin{equation}
\sqrt{\dfrac{(2J+1)(2J-3)}{2}}\rightarrow\sqrt{2J}
\end{equation}
\begin{equation}
PROD\rightarrow(4J)^{2m}=(8j)^{2m}
\end{equation}
Hence we have 
\begin{equation}
\dfrac{1}{\sqrt{PROD}}\rightarrow\dfrac{1}{(8j)^{m}}
\end{equation}

We use the Stirling approximation to calculate $FAC$. The detailed
results are given in Table~\ref{table:u9j}. 
\begin{table}[h]
\caption{ $\ln\dfrac{(C_{1}!)^{2}}{C_{2}!}\dfrac{C_{3}!}{C_{4}!C_{5}!}$ }

\label{table:u9j} \centering %
\begin{tabular}{r|r|r|r|r|r}
 & $-C_{i}$  & $\ln\sqrt{2\pi C_{i}}$  & $\alpha_{i}\ln C_{i}$  & $\gamma_{i}J\ln()$  & $\gamma_{i}J\ln()$ \tabularnewline
\hline 
(1)  & $\dfrac{}{}-2J$  & $2\ln(\sqrt{2\pi J})$  & $0$  & $2J\ln J$  & $0$ \tabularnewline
(2)  & $\dfrac{}{}+2J$  & $-\ln\sqrt{4\pi J}$  & $0$  & $-2J\ln(2J)$  & $0$ \tabularnewline
(3)  & $\dfrac{}{}-2J-\frac{1}{2}$  & $\ln\sqrt{8\pi J}$  & $\frac{1}{2}\ln(4J)$  & $2J\ln(4J)$  & $\frac{1}{2}$ \tabularnewline
(4)  & $\dfrac{}{}J+\frac{1}{2}$  & $-\frac{1}{2}\ln\sqrt{4\pi J}$  & $-\frac{1}{2}\ln(2J)$  & $-J\ln(2J)$  & $\frac{1}{2}$ \tabularnewline
(5)  & $\dfrac{}{}-J-\frac{1}{2}$  & $-\frac{1}{2}\ln\sqrt{4\pi J}$  & $\frac{1}{2}\ln(2J)$  & $-J\ln(2J)$  & $-\frac{1}{2}$ \tabularnewline
\hline 
Total  & $\dfrac{}{}-\frac{1}{2}$  & $\ln(\frac{\pi J}{2})^{1/4}$  & $\ln(2\sqrt{J})$  & $0$  & $\frac{1}{2}$\tabularnewline
\end{tabular}
\end{table}

\subsection{Combing Table 1 and Table 2}

There are many cancellations when we add the totals of $\ln FAC$
and $\ln3j$ in Table~\ref{table:3j} and Table~\ref{table:u9j}.
The result is 
\begin{equation}
\ln FAC+\ln3j=-(m-1)\ln2=-n\ln2
\end{equation}
The antilog is 
\begin{equation}
e^{-n\ln2}=\dfrac{1}{2^{n}}
\end{equation}
All the $j$ dependance 
 The $j$ dependence comes from 
\begin{equation}
\sqrt{\dfrac{(2J+1)(2J-3)}{2}}
\end{equation}
and PROD 
\begin{equation}
\sqrt{PROD}\rightarrow(8j)^{m}
\end{equation}
putting everything together we obtain the result: 
\begin{equation}
U9j\rightarrow\dfrac{(-1)^{n}}{2\sqrt{2}16^{n}}\dfrac{\sqrt{\left((2n+2)!(2n)!\right)}}{(n!)j^{n}}
\end{equation}

We note other work on asymptotics of CG coefficients by Reinsch and
Morehead \cite{Reinsch}. In their work they define

\begin{equation}
\beta=((j_{1}+j_{2}-j)(j+j_{2}-j_{1})(j+j_{1}-j_{2})(j_{1}+j_{2}+j))^{1/2}
\end{equation}
They find an approximate expression for the CG coeffecients in their
Eq.(B9).

\[
CG=\left<j_{1}j_{2}00|j0\right>\approx2(-1)^{\frac{j_{1}+j_{2}-j}{2}}\sqrt{\dfrac{2j+1}{2\pi\beta}}\sqrt{\dfrac{j+j_{1}+j_{2}}{j+j_{1}+j_{2}+1}}(1+\delta_{4}+\delta_{6})
\]

\begin{equation}
\times\left[1+\dfrac{1}{24}\left(\dfrac{2}{j}+\dfrac{2}{j_{1}}+\dfrac{1}{j_{2}}-\dfrac{1}{j+j_{1}+j_{2}}-\dfrac{1}{-j+j_{1}+j_{2}}-\dfrac{1}{j-j_{1}+j_{2}}-\dfrac{1}{j+j_{1}-j_{2}}\right)\right]
\end{equation}
We quickly run into trouble in making a comparison with our results,
especially for $n=0$. In their Eq.(B12) they have in the leading
term CG proportional to $\dfrac{1}{\sqrt{\beta}}$. However for the
case $j=j_{1}+j_{2}$, that is to say $I=I_{max}$, with our $n=0$,
we see that $\beta$ vanishes and hence their expression for CG blows
up. Evidently their formula is not valid in this region. On the other
hand our expression Eq.(13) from \cite{Kleszyk} works just fine.

%

\end{document}